# A 97% Peak Efficiency Single-Inductor-Multiple-Output DC-DC Converter with a Shared Bootstrap Gate Driver

Mohammadreza Zeinali
Electrical and Computer Engineering Dept., University of California, Los Angeles (UCLA), Los Angeles, CA, USA
Email: zeinali@ucla.edu

**Abstract**
This paper describes a SIMO DC-DC converter capable of generating all the required voltages (buck and boost) utilizing n-type transistors for the output switches. The design incorporates only one shared bootstrap capacitor and a single pad to connect the top plate of the off-chip bootstrap capacitor to the on-chip drivers. By utilizing one off-chip bootstrap capacitor and sharing it between the multiple outputs, this work presents a low cost and low area solution for implementing a SIMO DC-DC converter using n-type transistor switches. A prototype of the proposed SIMO DC-DC converter is fabricated in 180nm SILTERRA BCD Technology.

**Keywords:** SIMO, switching DC-DC converter, bootstrap gate driver, shared driver, high efficiency, high voltage driver.

## Introduction

Applications involving feature-rich system-on-chip (SoC) designs necessitate the use of multiple power supplies to enhance overall system performance and minimize power consumption. To achieve this, direct current (DC)-DC converters are commonly employed to convert a single power supply into multiple ones. Among the various converter types, switching LC DC-DC converters stand out, utilizing a capacitor and an inductor for each output regulated power supply. Despite their popularity, challenges such as electromagnetic interference, components costs, and physical size constraints limit the practicality of employing multiple inductors. A promising alternative lies in single-inductor-multiple-output (SIMO) DC-DC converters, particularly suitable for power management integrated circuits [1-4]. These converters offer the advantages of being both space- and cost-effective. Operation of such converters relies on charging the inductor current and discharging it into the output capacitor to regulate its voltage. As a result, a series of switches are involved, and the power efficiency of these converters is closely tied to the on-resistance of these switches. Consequently, minimizing the on-resistance of switches becomes a key objective. When considering a specific on-resistance, the preference for n-type transistor switches over p-type transistor switches arises due to the higher electron mobility in n-type transistors. However, the use of n-type transistors comes with the challenge of requiring a gate voltage higher than source voltage for turning on the switch, leading to the necessity of incorporating bootstrap gate drivers [1], [3]. An inherent drawback in employing a bootstrap gate driver lies in the need for a large off-chip capacitor to be charged in one phase, and subsequently applying the capacitor's voltage to the $V_{gs}$ of n-type transistor in the next phase. This, in turn, may demand the utilization of N off-chip bootstrap capacitors and 2N pads on a chip to regulate N output voltages, resulting in a design that is not only complex but also cost-intensive.

## Shared Bootstrap Gate Driver

This paper describes a SIMO DC-DC converter capable of generating all the required voltages (buck and boost) utilizing n-type transistors for the output switches. The design incorporates only one shared bootstrap capacitor and a single pad to connect the top plate of the off-chip bootstrap capacitor to the on-chip drivers. By utilizing one off-chip bootstrap capacitor and sharing it between the multiple outputs, this work presents a low cost and low area solution for implementing a SIMO DC-DC converter using n-type transistor switches.

Fig.1 shows the conventional SIMO DC-DC converter on the left and the proposed SIMO DC-DC converter using only a single shared bootstrap capacitor on the right. In the conventional design, each n-type output transistor uses its own large off-chip bootstrap capacitor which occupies 2 extra pads per output on the chip. This is while the proposed architecture uses a single shared bootstrap capacitor between the multiple outputs that requires only one extra pad on the chip for all outputs. For a SIMO DC-DC converter with N outputs, the proposed structure requires N-times lower off chip area and 2N-1 less pads on the chip compared to the conventional design.

Fig.2 shows the circuit details of the proposed SIMO DC-DC converter with only a single shared bootstrap capacitor. Each output has its own output delivery network Sout_j (including two n-type transistor switches Mn_L-j and Mn_R-j and one diode D_j) and its own driver network (including a High_Side_P-type_Driver_j driving a p-type transistor Mp_j and a High_Side_N-type_Driver_j driving a n-type transistor Mn_g_j). Two transistors Mn_L-j and Mn_R-j are placed in a manner to have back-to-back source to drain diodes to guarantee that the current passes only through the on-resistance of the switches. This is because the body of these transistors are not accessible in this technology. The single shared off-chip bootstrap capacitor shares one pad with the VL+ of off-chip inductor where its bottom plate is connected and just a single extra pad is required to connect its top plate to the on-chip drivers. The controller accepts the outputs as the input and delivers a control signal per output for controlling the output drivers. The operation of the proposed driver is as follows:

1) By turning on Sp and Sy switches, inductor current will be charged based on $V_{in} = L\frac{di}{dt}$. It is worth mentioning that in this phase, the bottom plate of bootstrap capacitor is also connected to gnd. Consequently, the bootstrap capacitor is charged to Vc through diode $D_B$ which is the required voltage to turn on output switches correctly.

2) In the next phase one of the output switches should be turned on to discharge the inductor current to the output capacitor and regulate its voltage. The interesting point is that when the output switch is turned on (i.e., $S_{out\,j}$) and Sy switch is off, the three nodes $V_{L+}$, $V_{Sj}$, and $V_{outj}$ are equal and are connected to the bottom plate of the bootstrap capacitor. As a result, the top plate of the bootstrap capacitor will be equal to $V_{out\,j}+V_C$ which will be connected to the gate of the output switch through $M_{pj}$ switch.

The detailed implementation of a single output driver network is shown in Fig.3. The VSG voltage of the p-type driver transistor is provided through the resistor divider formed by R1 and R2 and enabled by the bottom n-type transistor Mn_5_j. Also, the four stacked diode-connected n-type transistors Mn_1_j to Mn_4_j connected in parallel to R1 limit the source gate voltage of the p-type driver transistor (to the value of four times of the gate-source voltage of n-type transistors Mn_1_j to Mn_4_j).

### Measurement Results

Fig.4 illustrates the operation of proposed SIMO converter to generate four output voltages and waveforms of critical nodes including inductor current (IL), VL+, and VC+, where from left to right, 1st on-chip output produces 12V, then the 1st off-chip output provides 10V and then the 2nd off-chip output generates 3.3V and finally the 2nd on-chip output provides 1.8V. The top plate voltage of the shared bootstrap capacitor stands VC=5V higher than the selected output while in the idle time it keeps at 5V. When the selected output is ready the triangle-shaped inductor current peaks about 135 mA. The ripple of the measured on-chip 12V and 1.8V outputs are demonstrated respectively in Fig.5 which is lower than 20 mV.

Fig. 6 shows a comparison between this work and the State-of-the-Art SIMO converters based on the number of bootstrap capacitors and pads used to drive the output switches. It is obvious that n-type switches can provide better efficiency due to lower on-resistance in the same switch area.

A prototype of the proposed SIMO DC-DC converter with two on-chip outputs and two off-chip outputs is fabricated in 180nm SILTERRA BCD Technology for which the chip micrograph is shown above. Two output driver and delivery networks are implemented on-chip and the other two are implemented off-chip due to limited chip area while all four outputs share the same single off-chip bootstrap capacitor.

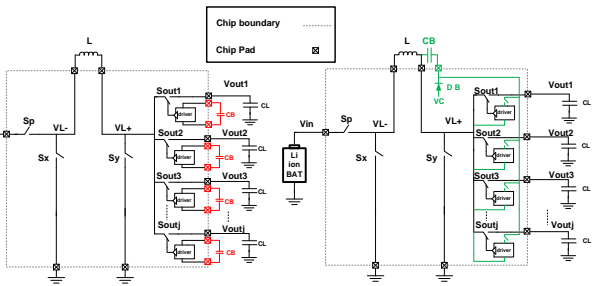

**Fig.1** Structure of conventional SIMO converter with N bootstrap capacitors and 2N pads for N output supply voltages (left), and proposed SIMO converter with a shared bootstrap capacitor and only one pad to drive N output switches (right).

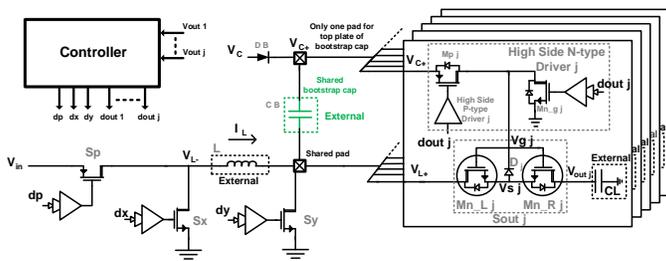

**Fig.2** Proposed SIMO DC-DC converter with shared bootstrap capacitor driver.

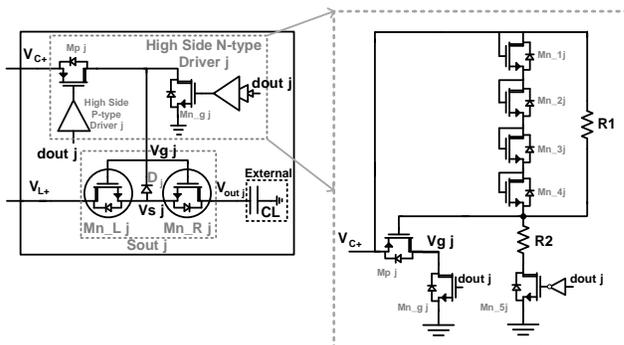

**Fig.3** Schematic of the high side N-type driver and P-type driver.

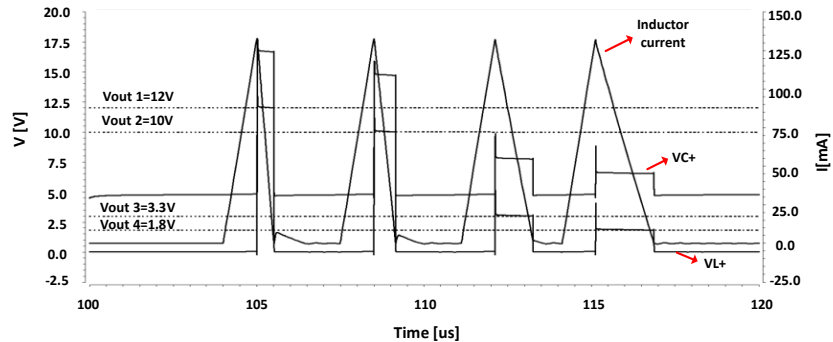

**Fig.4** Waveforms of critical nodes (VL+, VC+, and Indictor current).

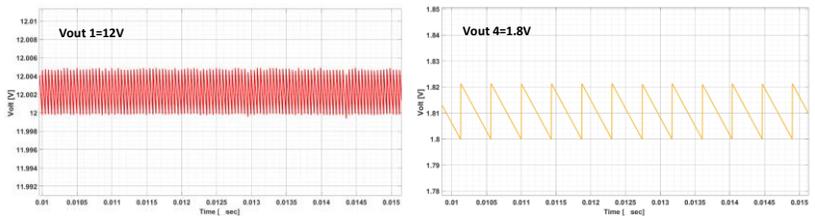

**Fig.5** Waveforms and ripples of two output voltages generated using on-chip switches.

| | | ISSCC' 2015 [1] | ISSCC' 2015 [2] | ISSCC' 2018 [3] | ISSCC' 2023 [4] | This work |
|---|---|---|---|---|---|---|
| | Input Voltage (V) | 5 | 3.6~5 | 4~40 | 1.8 | 1.5~4.2 |
| | Number of Outputs (Range) | 10 (1.8/2.5/2 .8/3.3) | 4 (1.2~3.3) | 3 (1.8/4.5/2 1) | 4 (0.8~1.4) | 4 (1~12) |
| | Max. Efficiency | 88.7% | 90 % | 86 % | 87.3 % | 97 % |
| Gate Drive Cost | Bootstrap capacitors | 0-PMOS Switches | 4 | 3 | 0-PMOS Switches | 1 |
| | Number of Pads for bootstrap capacitor | 0 | 8 | 6 | 0 | 1 |
| | Switch area (mm2) | 0.44 | 0.15 | 0.12 | 0.14 | 0.07 |
| | Ripple (mV) | <40 | <17 | <25 | <20 | <20 |
| | Switching Frequency | 0.7 MHz | 1 MHz | 2.4 MHz | 10 MHz | 0.2-10 MHz |
| | Technology | 350 nm | 250 nm | 110 nm | 16nm | 180 nm |

**Fig.6** Comparison of this work with the recent State-of-the-Art SIMO converters.

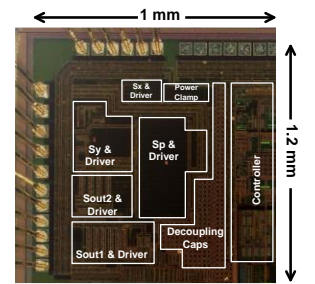

**Fig.7** Chip micrograph of the proposed SIMO converter.